\documentclass[fleqn,12pt,twoside]{article}
\usepackage[headings]{espcrc1}

\readRCS
$Id: sorensen\_qm2005.tex,v 1.2 2004/02/24 11:22:11 spepping Exp $
\usepackage{graphicx}
\usepackage[figuresright]{rotating}

\newcommand{\AmS}{{\protect\the\textfont2
  A\kern-.1667em\lower.5ex\hbox{M}\kern-.125emS}}

\title{Elliptic flow fluctuations in Au+Au collisions at
  $\sqrt{s_{_{NN}}}=200$~GeV }

\author{P. Sorensen\address[MCSD]{Brookhaven National Laboratory, 
    Upton, New York 11973-5000, USA} for the STAR Collaboration\thanks{For the full list of STAR authors and acknowledgements, see appendix `Collaborations' of this volume} }%
       
\runtitle{$v_2$ fluctuations}
\runauthor{P. Sorensen}

\begin{document}
\maketitle

\begin{abstract}
  In this talk, we report an analysis of elliptic flow ($v_2$), non-flow
  ($\delta_{2}$), and $v_2$ fluctuations ($\sigma_{v_{2}}$) for the
  STAR collaboration at middle rapidity in $\sqrt{s_{_{NN}}}=200$~GeV
  Au+Au collisions. We compare the results to models of the
  fluctuations in the initial eccentricity of the overlap zone.
\end{abstract}

\textit{Introduction:} Understanding the early-time dynamics in
heavy-ion collisions is an important step in understanding the
properties of the matter created in the collision overlap zone.
Elliptic flow ($v_2$) measurements are sensitive to the shape of the
initial overlap zone~\cite{v2papers} so $v_2$ fluctuations can reveal
information about fluctuations and correlations in the initial
geometry. Those fluctuations and correlations will lead to a better
understanding of the initial conditions of the collision evolution.
We report on a study of $v_2$ fluctuations based on 15 million events
measured with the STAR detector~\cite{STAR}. A similar analysis has
been carried out by PHOBOS~\cite{phobos} and results have been
discussed in the literature~\cite{flowfl}. Please note, data shown in
this talk, showing small $v_2$ for events containing jets, was in error.

\textit{Analysis method:} The q-distribution method can be used to gain access to information about azimuthal correlations and $v_2$ fluctuations. 
The $n^{th}$ harmonic reduced flow vector is
defined as $q_{n,x} = \frac{1}{\sqrt{M}} \Sigma_i\cos(n\phi_i)$ and
$q_{n,y} = \frac{1}{\sqrt{M}} \Sigma_i\sin(n\phi_i)$, where $M$ is the
number of tracks and $\phi_i$ is the azimuth angle of a track with
respect to a preferred emission axis~\cite{qdist}. The sums over the x and y components of the particle momentum are equivalent to random walks so that the central limit theorem is applicable for large $M$, in which case the
$\vec{q_{n}}$ distribution will be a 2-D Gaussian with widths that depend on particle correlations and $v_2$ fluctuations. According to the
definition of $v_2$, if the reaction-plane is the preferred axis, the
2-D Gaussian will be shifted along $x$ by $\sqrt{M}v_n$ with widths:
$\sigma_x^2 \simeq \frac{1}{2}\left(1 + M\delta_n\right)$
and $\sigma_y^2 \simeq \frac{1}{2}\left(1 + M\delta_n\right)$.
$\delta_n=\langle\cos(n\Delta\phi)\rangle - v_n^2$ are correlations
unrelated to the reaction plane~\cite{qdist}. Neither the preferred
axis, nor the reaction plane direction are known on an event-by-event
basis, so we calculate the magnitude $|q_n|$. The 2-D Gaussian then
becomes:
\begin{equation}
\frac{dN}{q_ndq_n} = 
\frac{exp\left(-\frac{q_n^2+Mv_n^2\{q\}}{2\sigma_x^2}\right)}{\sqrt{\pi}\sigma_x\sigma_y}
\sum_{k=0,2,...}^{\infty}\left(1-\frac{\sigma_x^2}{\sigma_y^2}\right)^k
\left(\frac{q_n}{v_n\{q\}\sqrt{M}}\right)^k
\frac{1}{k!}
\Gamma\left({\scriptstyle \frac{2k+1}{2}}\right)
I_k\left({\scriptstyle \frac{q_nv_n\{q\}\sqrt{M}}{\sigma_x^2}}\right) 
\label{dq}
\end{equation}
where $\Gamma$ and $I_k$ are gamma and modified Bessel's functions. If
we assume the preferred axis for the $dN/d|q|$ is the reaction-plane,
then $v_2\{q\}=v_2$ and $v_2$ fluctuations can be introduced by
convoluting the above distribution with a $v_2$ distribution
($dN/dv_{2}$). This introduces an added width to $dn/d|q_{2}|$:
$\sigma_x^2 \simeq \frac{1}{2}\left(1 + M\delta_2 + 2M\sigma_{v_{2}}^2
\right)$, where $\sigma_{v_{2}}$ is the r.m.s. width of
$dN/dv_{2}$. $\delta_2$ and $\sigma_{v_{2}}^2$ contribute to the width
with a factor of $M$, so it is not possible to extract these
quantities independently as previously
thought~\cite{Sorensen:2006nw}. Rather, we define the dynamic width of
$dN/d|q_{2}|$ as $\sigma_{dyn}^2=\delta_{2}+2\sigma_{v_{2}}^2$, which
can be readily extracted from data.

We don't know a priori what the preferred emission axis is. For
peripheral Au+Au or p+p collisions, the preferred axis may be
predominantly determined by the thrust axis of outgoing
partons. However, for mid-central and central collisions, the four-particle cumulant $v_2\{4\}$ (a higher multiplicity correlation analysis~\cite{Adler:2002pu})
decreases with eccentricity suggesting the geometry of the overlap
determines the preferred axis.

Distinguishing between $\sigma_{v_2}^2$ and $\delta_2$ requires
knowledge of the reaction-plane or information about the
$\Delta\phi$-, $\Delta\eta$-, charge-sign- or multiplicity-dependence
of non-flow. STAR addresses this in two general ways, 1) analyzing
higher multiplicity correlations as in a q-distribution or higher
order cumulant analysis~\cite{Adler:2002pu,v2papers} or 2) analyzing
the $\Delta\phi$, $\Delta\eta$, or charge-sign dependence of
correlations. Examples of 2) include measuring two-particle
correlations across a wide $\eta$-gap~\cite{Voloshin:2007af},
correlating only like-sign particles, or by fitting the $\Delta\phi$,
$\Delta\eta$ correlation density to isolate the $\cos2(\Delta\phi)$
term $v_2\{2D\}$~\cite{Trainor:2007fu}. But the small $\Delta\phi$
correlations in Au+Au collisions stretch over a large $\Delta\eta$
range~\cite{ridge} and it's possible that the space-momentum
correlation inducing mechanism responsible for $v_2$ may contribute to
the near-side correlation by translating quantum- or
geometry-fluctuations from the initial overlap to correlations in
momentum-space~\cite{radialflow,Dumitru} linking the ridge and
$v_2$~\cite{Hwa:2008um}.

For $dN/dv_{2}$, we consider a Gaussian and Bessel-Gaussian
(BG($v_{BG},\sigma_{v_{BG}}$))~\cite{planes}. Convoluted with
Eq.~\ref{dq}, both give an equally good description of $dN/d|q|$. The
mean and r.m.s. of the BG or the parameters $v_{BG}$ and
$\sigma_{v_{BG}}$ can be reported. We find that the BG mean and
r.m.s. are within errors the same as the Gaussian mean and r.m.s. The
use of a BG shape for the $dN/dv_{2}$ is motivated by the shape of the
participant eccentricity~\cite{epart} distribution in a Glauber Monte
Carlo model~\cite{Miller:2003kd}.
In the case that $v_2 \propto \varepsilon_2$, the parameters extracted
from the BG are either related to the participant-plane or the
reaction-plane. The parameter $v_{BG}$ represents the $v_{2}$ computed
with respect to the reaction-plane, while $\langle v_{2} \rangle$
represents $v_2$ computed with respect to the
participant-plane~\cite{planes}. Here, we report the mean and rms
width of a Gaussian and compare it to models of $\varepsilon_{part}$.

\begin{figure}[htb]
  \hspace{1cm}
  \resizebox{0.425\textwidth}{!}{\includegraphics{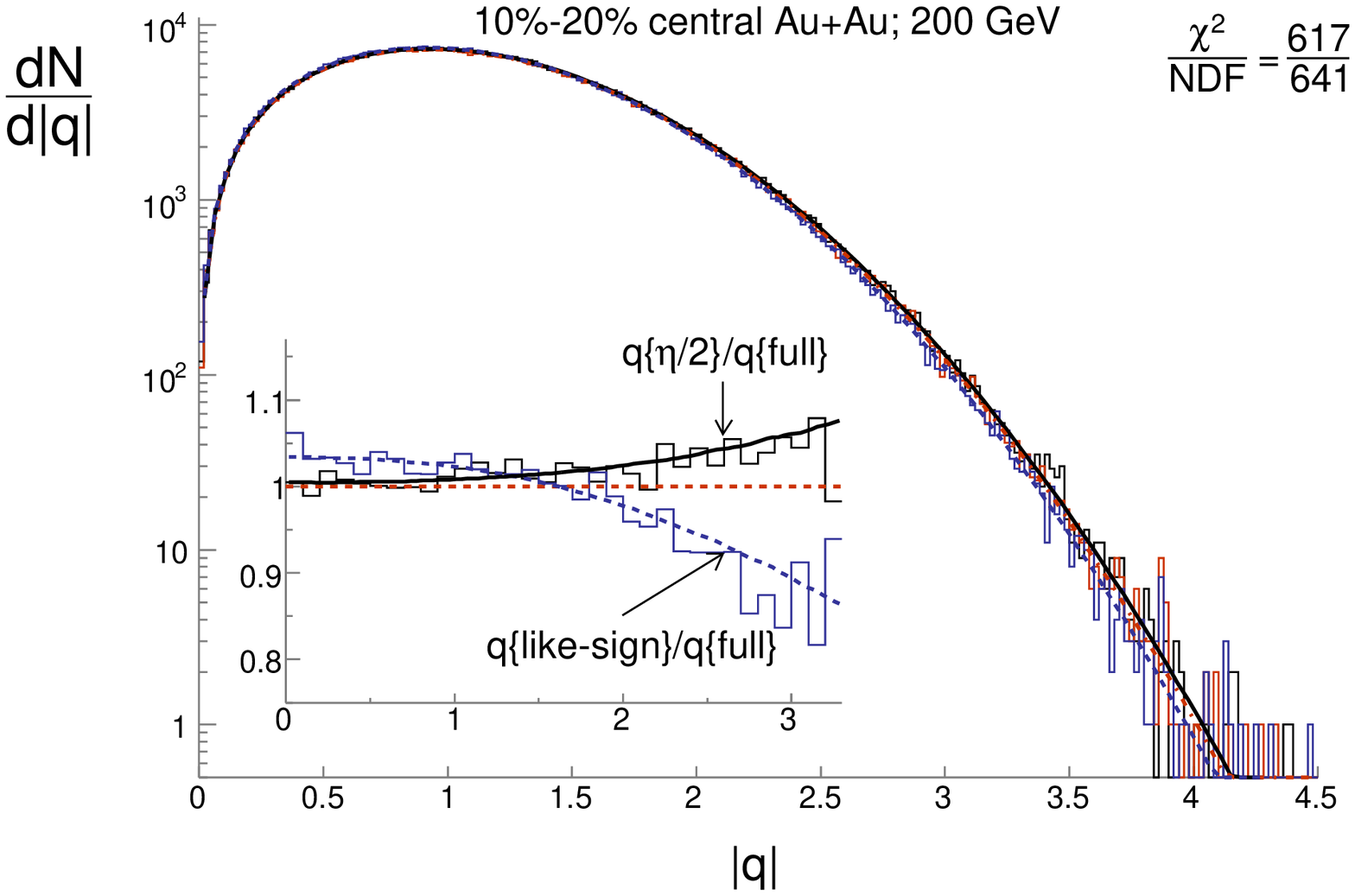}}
  \resizebox{0.425\textwidth}{!}{\includegraphics{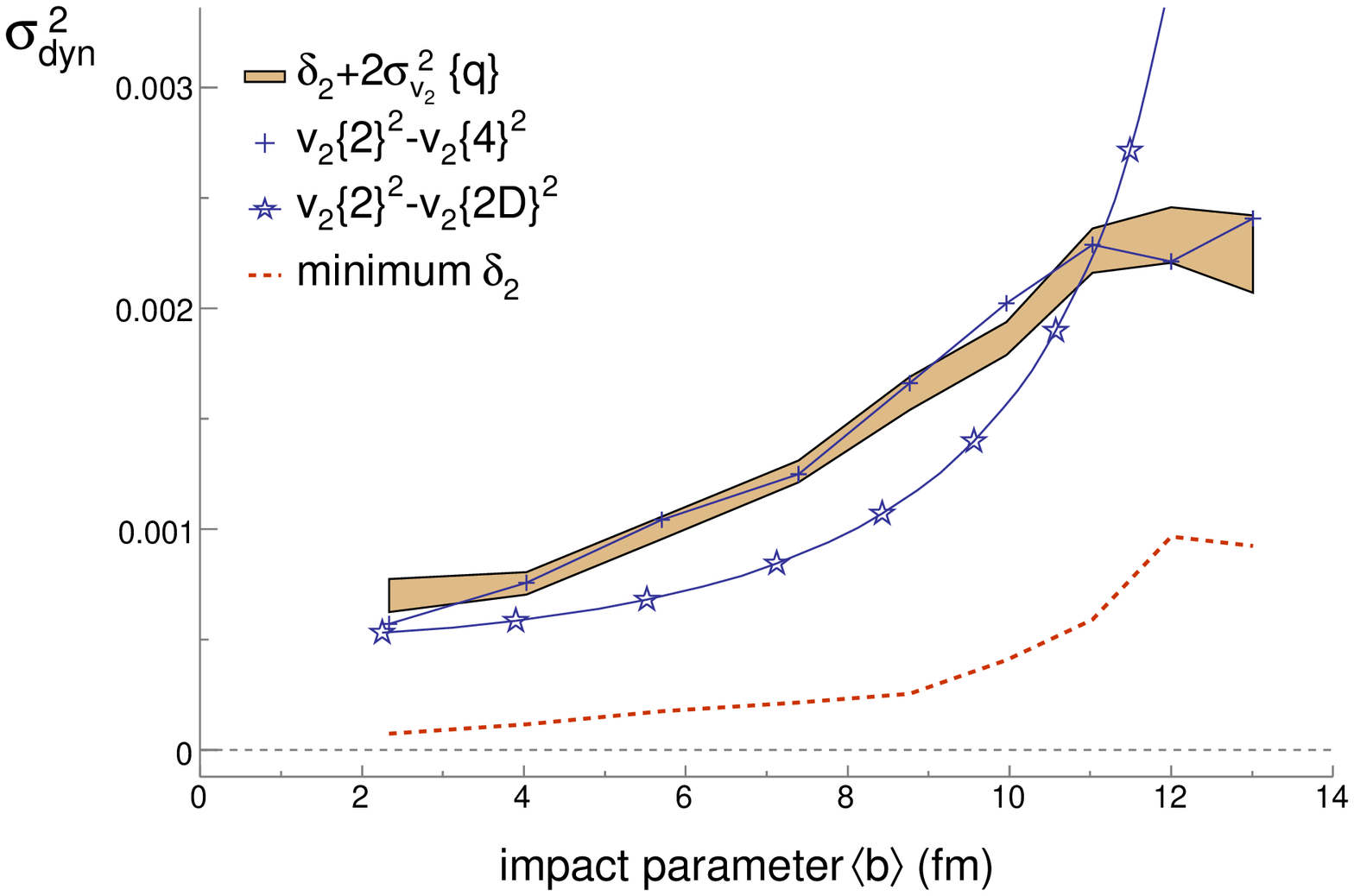}}
  \vspace{-1.0cm}
  \caption[]{ Left: $dN/d|q|$ for like-sign tracks (q\{like-sign\}),
    tracks within one hemisphere of the STAR acceptance
    (q\{$\eta/2$\}), and all tracks with $0.15 < p_T < 2.0$~GeV/c
    (q\{full\}). The inset shows the ratio of each distribution to
    q\{full\}. Right: The dynamic width of $dN/d|q|$ with width from
    cumulant analysis. $v_2\{2\}^2-v_2\{2D\}^2$ gives the non-flow
    $\delta_{2}$ inferred from a fit to the 2D correlation
    densities~\cite{Trainor:2007fu}.  }
\label{f1}
\end{figure}

\textit{Results:} Fig.~\ref{f1} (left) shows our fits to $dN/d|q_2|$
where $|q_2|$ is calculated from three track samples: full-acceptance
(q\{full\}), reduced-acceptance (q\{$\eta/2$\}), and full-acceptance
like-sign (q\{like-sign\}). The different samples are expected to
contain different contributions of correlations; HBT,
resonance decays, and parton fragments, which contribute to non-zero values of $\delta_2$. The inset shows the ratio of
each distribution to $dN/d|\mathrm{q\{full\}}|$. The widths are
ordered as follows: $\sigma_{dyn}\{\mathrm{q\{}\eta/2\mathrm{\}}\} >
\sigma_{dyn}\{\mathrm{q\{full\}}\} >
\sigma_{dyn}\{q\mathrm{\{like-sign\}}\}$. This is expected because the
like-sign sample removes most correlations caused by resonance
decay. Jet fragments exhibit charge-ordering so those correlations are
also reduced. The $\eta/2$ sample gives a greater weight to the
short-range correlations so $\delta_2$ is enhanced.

\vspace{-0.0cm}

\begin{figure}[htb]
  \hspace{1cm}
  \resizebox{0.425\textwidth}{!}{\includegraphics{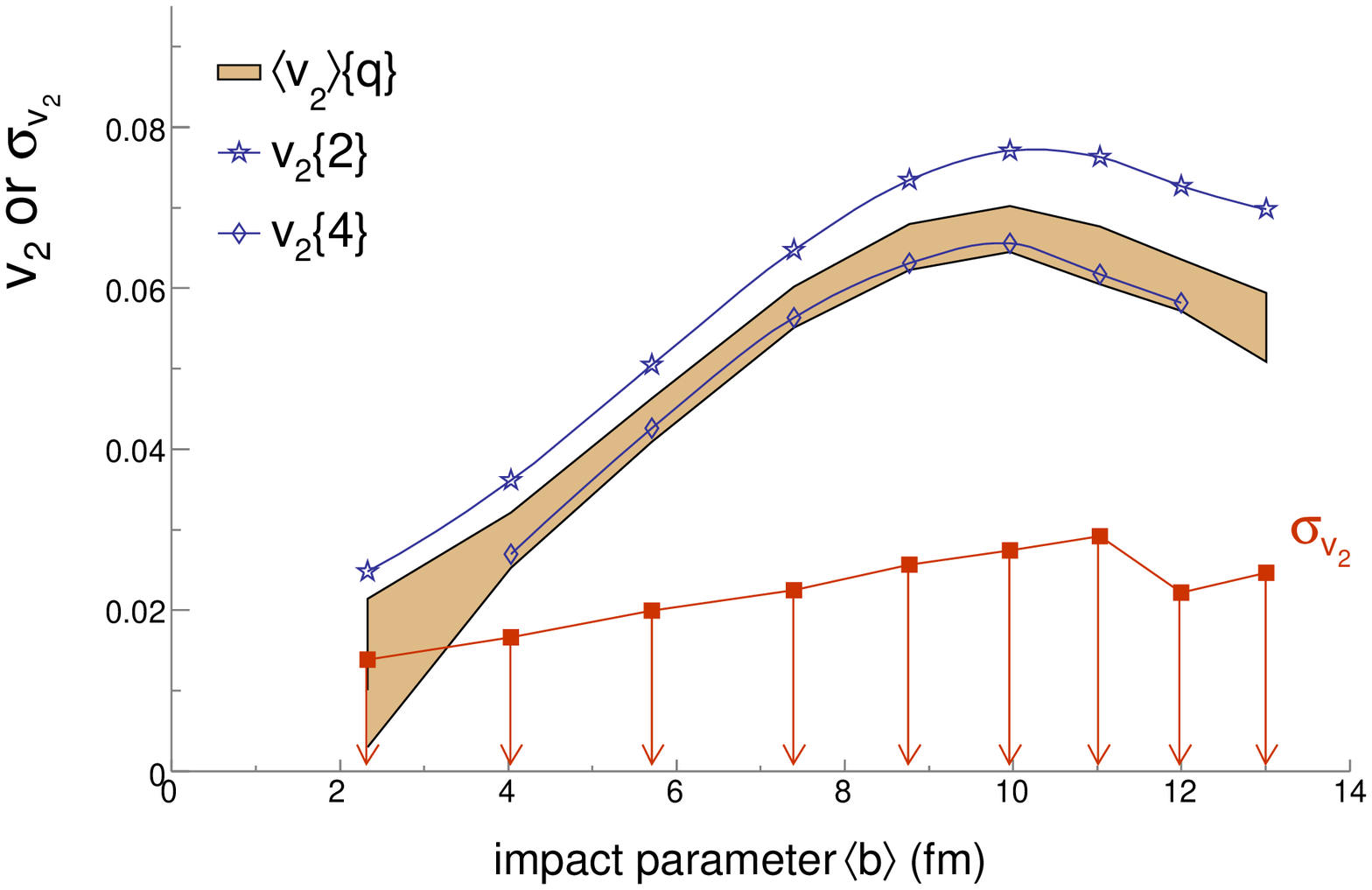}}
 \resizebox{0.425\textwidth}{!}{\includegraphics{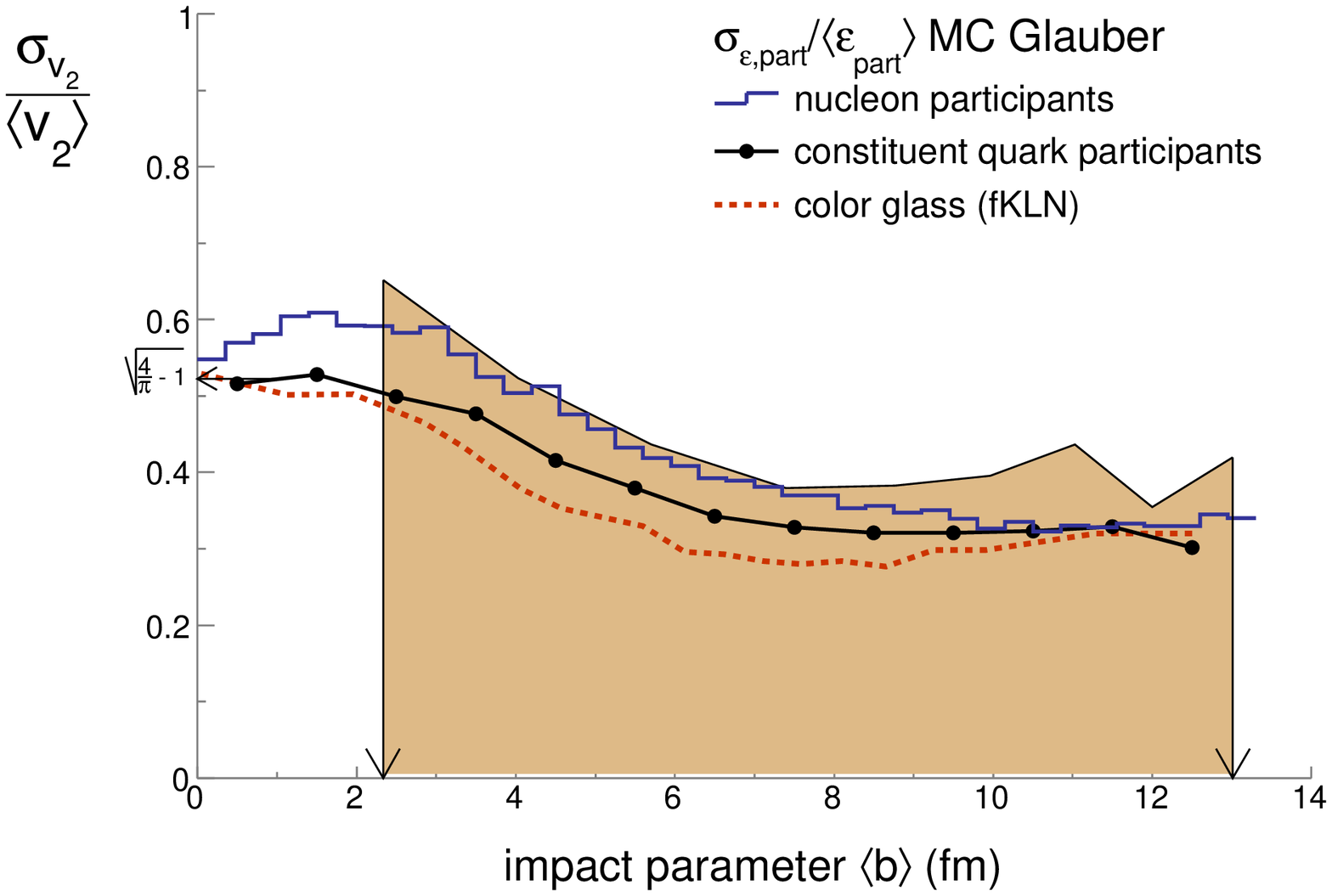}}
  \vspace{-1cm}
 \caption[]{ Left: The tan band shows the allowed values of $\langle
   v_2\rangle$ from this analysis and the upper limit on
   $\sigma_{v_{2}}$. Right: Upper limit on $\sigma_{v_{2}}/\langle
   v_{2}\rangle$ compared to models of eccentricity. }
\label{f2}
\end{figure}

Fig.~\ref{f1} (right) shows the dynamic width of the full-acceptance
$dN/d|q|$. The tan band includes systematic uncertainties from
acceptance, finite multiplicity and comparisons of methods including
the dynamic width from the cumulants $v_2\{2\}^2 - v_2\{4\}^2 =
\delta_2 + 2\sigma_{v_{2}}^2$. It was shown in ref.~\cite{planes} that
within the precsion of our measurements we cannot extract more
information from $dN/d|q|$ than that already found from the
cumulants. In particular, if the $v_2$ fluctuations are Gaussian, then
the equality $v_2\{2\}^2 - v_2\{4\}^2 = \delta_2 + 2\sigma_{v_{2}}^2$
is exact. Published STAR $v_2\{2\}$ and $v_2\{4\}$ values
vs. centrality have been calculated by weighting each event by the
number of tracks in the event~\cite{v2papers}. Those are not
comparable to the q-distribution analysis where each event is weighted
equally. We therefore compare to cumulant data without the
multiplicity weighting and find good agreement. The minimum value of
$\delta_2$ shown in Fig.~\ref{f1} (right) is calculated from the
difference between $\langle q*q\rangle$ calculated with same-charge
pairs and unlike-sign pairs. We also show $\delta_{2}\{2D\}=
v_2\{2\}^2 - v_2\{2D\}^2$ which shows that a large component of
$\sigma_{dyn}^2$ is accounted for by the near-side peak in the 2D
correlations~\cite{Trainor:2007fu}. Fig.~\ref{f2} (left) shows the
allowed values for $\langle v_2\rangle$ and $\sigma_{v_{2}}$. The case
of zero fluctuations is not excluded and an upper limit is reported.

\vspace{-0.0cm}


Fig.~\ref{f2} (right) shows $\sigma_{v_2}/ \langle v_2\rangle$
calculated using a Gaussian for $dN/dv_2$. The upper limit on this
ratio is compared to several Monte Carlo models of
$\varepsilon_{part}$ fluctuations. In the nucleon Glauber models, the
eccentricity $\varepsilon = \langle y^2 - x^2\rangle / \langle y^2 +
x^2\rangle$ is calculated based on the $x$ and $y$ cooridinates of
participant nucleons. For the constituent quark Glauber model, each
nucleon is treated as three constituent quarks bound within the
nucleon, and the eccentricity is calculated based on the constituent
quarks that participate in the collision. All models lie within the
allowed range while the CGC model~\cite{cgc} and the model based on
constituent quarks both have smaller
relative widths. The nucleon participant model leaves little room for
other sources of fluctuations and correlations beyond the initial
geometric ones. The large near-side peak observed in two-particle
correlations contradict the idea that all or most of the width of
$\sigma_{dyn}^2$ is dominated by $\sigma_{v_2}$ suggesting that the
CGC or constituent-quark model may be preferred. Recently it was
proposed that correlations and fluctuations in the initial conditions
may also contribute to the near-side ridge~\cite{Dumitru}.


\textit{Conclusions:} The dynamic width of $dN/dq$ has been
presented. The width reflects correlations and $v_2$ fluctuations. An
upper limit on $\sigma_{v_{2}}$ has been presented and compared to
models of $\varepsilon_{part}$ fluctuations. From our comparison to the
Glauber MC we conclude that either the density fluctuations in the
initial state are manifested in both the near-side peak and in
$\sigma_{v_{2}}$, or the initial overlap density is smoother than would
be expected from a Glauber model.

\end{document}